\documentclass[a4paper,11pt]{article}
\pdfoutput=1

\usepackage{jcappub}

\usepackage[T1]{fontenc}
\usepackage[separate-uncertainty = true]{siunitx}

\title{Measuring the Polarization Reconstruction Resolution of the ARIANNA Neutrino Detector with Cosmic Rays}

\author[a]{A.~Anker}
\author[b]{P. Baldi}
\author[a]{S. W. Barwick}
\author[c]{J. Beise}
\author[d,e]{D. Z. Besson}
\author[f]{S. Bouma}
\author[f]{M. Cataldo}
\author[g]{P. Chen}
\author[a]{G. Gaswint}
\author[c]{C. Glaser}
\author[c]{A. Hallgren}
\author[h]{S. Hallmann}
\author[i]{J. C. Hanson}
\author[j]{S. R. Klein}
\author[k]{S. A. Kleinfelder}
\author[f]{R. Lahmann}
\author[a]{J. Liu}
\author[d]{M. Magnuson}
\author[b]{S. McAleer}
\author[h,f]{Z. S. Meyers}
\author[g]{J. Nam}
\author[f,h]{A. Nelles}
\author[d,e]{A. Novikov}
\author[a]{M. P. Paul}
\author[a]{C. Persichilli}
\author[f,h]{I. Plaisier}
\author[h,f]{L. Pyras}
\author[a]{R. Rice-Smith}
\author[l]{J. Tatar}
\author[g]{S.-H Wang}
\author[f,h]{C. Welling}
\author[a,1]{L. Zhao \note{Corresponding author}}

\affiliation[a]{Department of Physics and Astronomy, University of California, Irvine, CA 92697, USA.}
\affiliation[b]{Department of Information and Computer Science, University of California, Irvine, CA 92697, USA.}
\affiliation[c]{Uppsala University Department of Physics and Astronomy, Uppsala SE-752 37, Sweden.}
\affiliation[d]{Department of Physics and Astronomy, University of Kansas, Lawrence, KS 66045, USA.}
\affiliation[e]{National Research Nuclear University MEPhI (Moscow Engineering Physics Institute), Moscow 115409, Russia.}
\affiliation[f]{ECAP, Friedrich-Alexander Universität Erlangen-Nürnberg, 91058 Erlangen, Germany.}
\affiliation[g]{Department of Physics and Leung Center for Cosmology and Particle Astrophysics, National Taiwan University, Taipei 10617, Taiwan.}
\affiliation[h]{DESY, 15738 Zeuthen, Germany. }
\affiliation[i]{Whittier College Department of Physics, Whittier, CA 90602, USA.}
\affiliation[j]{Lawrence Berkeley National Laboratory, Berkeley, CA 94720, USA.}
\affiliation[k]{Department of Electrical Engineering and Computer Science, University of California, Irvine, CA 92697, USA.}
\affiliation[l]{Research Cyberinfrastructure Center, University of California, Irvine, CA 92697, USA. }

\emailAdd{leshanz@uci.edu, sbarwick@uci.edu, christian.glaser@physics.uu.se}

\abstract{The ARIANNA detector is designed to detect neutrinos with energies above \SI{e17}{eV}. Due to the similarities in generated radio signals, cosmic rays are often used as test beams for neutrino detectors. Some ARIANNA detector stations are equipped with antennas capable of detecting air showers. Since the radio emission properties of air showers are well understood, and the polarization of the radio signal can be predicted from the arrival direction, cosmic rays can be used as a proxy to assess the reconstruction capabilities of the ARIANNA neutrino detector. We report on dedicated efforts of reconstructing the polarization of cosmic-ray radio pulses. After correcting for difference in hardware, the two stations used in this study showed similar performance in terms of event rate and agreed with simulation. Subselecting high quality cosmic rays, the polarizations of these cosmic rays were reconstructed with a resolution of \ang{2.5} (68\% containment), which agrees with the expected value obtained from simulation. A large fraction of this resolution originates from uncertainties in the predicted polarization because of the contribution of the subdominant Askaryan effect in addition to the dominant geomagnetic emission. Subselecting events with a zenith angle greater than \ang{70} removes most influence of the Askaryan emission, and, with limited statistics, we found the polarization uncertainty is reduced to \ang{1.3} (68\% containment).
}

\collaboration{ARIANNA collaboration}

\begin{document}
\maketitle
\flushbottom

\section{Introduction}
The ARIANNA experiment \cite{COSPAR2019} searches for radio flashes from ultra-high-energy (UHE, $E_\nu > \SI{e17}{eV}$) neutrinos interacting in polar ice sheets. In-ice radio detection is considered a cost-efficient technique to instrument the large volumes required to measure the small flux of UHE neutrinos. The ARIANNA experiment explores the technical feasibility of the technique and has been providing guidance for RNO-G \cite{RNOGWhitePaper2021} and the future IceCube-Gen2 \cite{Gen2WhitePaper}.
In a uniquely radio quiet area on the Ross-Ice-Shelf in Antarctic, a hexagonal array of pilot-stations has been taking data for several years. In addition, two detector stations have been installed at the South Pole. In previous work, a limit on the high-energy neutrino flux has been derived \cite{ARIANNALimit2020}. The sensitivity of the detector was optimized through optimizations of the signal chain \cite{Glaser2020Bandwidth} and trigger \cite{ICRC2021Trigger}, and the reconstruction abilities have been investigated. 

The reconstruction of the direction and energy of the neutrino requires the measurement of the distance to the neutrino vertex, the signal arrival direction, the viewing angle, and the signal polarization \cite{GlaserICRC2019}. A novel method to determine the vertex distance was developed and tested in an in-situ measurement on the Ross-Ice-Shelf \cite{DnR2019}. The ability to measure the signal direction and polarization was determined in an in-situ measurement at the South Pole \cite{COSPAR2019, GaswintICRC2019, ARIANNA2020Polarization}. The ability to measure the viewing angle and to combine all individual measurements to estimate the neutrino direction and energy was quantified in a simulation study using the forward folding technique \cite{NuRadioReco,GaswintPhD, ARIANNADirectionICRC2021} as well as through deep neural networks \cite{ICRC2021DLenergy, ICRC2021DLdirection}. 

The measurement of radio signals from cosmic-ray air showers provides another opportunity to study the detector capabilities under realistic conditions, as detailed in this paper. 
In addition to downward-facing log-periodic dipole antennas (LPDAs) and dipole antennas designed for in-ice shower detection, some of the detector stations are equipped with upward-facing LPDAs optimized for air shower detection. As the spacing between stations is so large, air-shower characteristics have to be reconstructed from a single station as opposed to how typical dedicated air-shower radio arrays perform such measurements\cite{Barwick2017, NellesICRC2019, Welling2019}.

For neutrino signals, the polarization direction is important because it is needed to reconstruct the neutrino arrival direction \cite{GlaserICRC2019, GaswintPhD, ARIANNADirectionICRC2021,RNO-G:2021zfm}. In previous studies we have shown that polarization is the dominant uncertainty on the neutrino direction \cite{GlaserICRC2019,GaswintPhD,ARIANNADirectionICRC2021}. 
The other parameters that impact the neutrino direction, namely signal arrival direction \cite{Kim:2020qyd} and viewing angle, can typically be measured more precisely. 
Thus, an accurate reconstruction of signal polarization is the key to improve reconstructed neutrino angular resolution.

The frequency content and duration of radio pulses generated by cosmic rays are similar to what is expected for neutrinos \cite{Schoorlemmer:2015afa,Welling:2021cgl}. Also, since the ARIANNA LPDAs are buried in snow, the local effects on antenna response for cosmic-ray detection are similar to the local effects for neutrino detection. This makes the polarization measurement of cosmic rays a useful tool to certify the expected performance of the ARIANNA stations. 

Another useful feature is that the radio emission from air showers is well understood \cite{Huege2016}. The air-shower radio signal is dominated by the geomagnetic emission whose polarization can be calculated based on the air-shower direction and the geomagnetic field \cite{Scholten:2016gmj}. As a consequence, the cosmic-ray polarization is predominantly horizontal for most cosmic-ray directions \cite{AERAPolarization,Schellart:2014oaa}. In addition to the geomagnetic emission component, the Askaryan effect, i.e., a time-varying negative charge-excess in the shower front, contributes to the air-shower radio signal \cite{Werner2007, Vries2010a}. The Askaryan component is radially polarized towards the shower axis and thus depends on the relative position of the observer to the shower axis. The relative contribution is at the 10\% level \cite{AERAPolarization,Schellart:2014oaa} and decreases further with increasing zenith angle \cite{GlaserErad2016} for the magnetic field at Ross-Ice-Shelf. Therefore, the polarization can be approximated with the geomagnetic emission and predicted with good precision from the air-shower direction, whereas the polarization direction of neutrinos is unknown a priori.  This unique feature facilitates reconstruction.

\section{Overview of data and detector stations}

Data from two detector stations (station 32 and station 52) capable of air-shower detection were used in this study, comprising the two time periods Dec 1 2017 - Mar 15 2018 and Dec 1 2018 - Mar 15 2019. 

We first focus on data from the 2018-2019 season, which better represents the designed working environment of the stations.  In the 2018-2019 season, the stations were buried deeper in the snow due to a year of additional snow accumulation compared to 2017-2018. With the extra depth of snow, the actual signal propagation better matches the signal propagation model, which assumes antennas surrounded by uniform, infinite snow.

\begin{figure}[t]
\centering
  \includegraphics[scale=1]{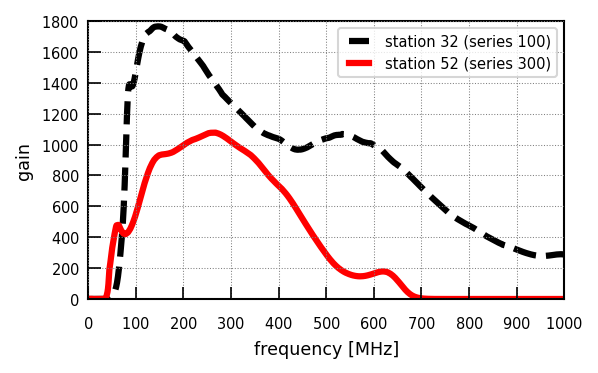}
  \caption{Gain as a function of frequency for the amplifier models in the two different ARIANNA stations.}
  \label{fig:amp_gains}
\end{figure}

The two detector stations were equipped with different hardware configurations. Station 52 is equipped with four upward-facing LPDAs, two downward-facing LPDAs, and two dipole antennas. The four upward-facing LPDAs consist of two pairs of parallel LPDAs with the planes of the two pairs orthogonal to each other.  Station 32 is equipped with the same geometry of four upward-facing LPDAs, but it is not equipped with either downward-facing LPDAs or dipole antennas. Whereas station 52 is equipped with a series-300 amplifier, station 32 employs a series-100 amplifier; these amplifiers exhibit different gain characteristics, as a function of frequency, as shown in Fig.~\ref{fig:amp_gains}. The difference in gains results in different thermal noise levels for the two stations. Station 32 has a post-amp RMS thermal noise of \SI{20}{mV}, while station 52 has a post-amp RMS thermal noise of \SI{10}{mV}. The trigger for both stations requires the voltage waveform to exceed a high and a low threshold within \SI{5}{ns} on one upward-facing channel, and again on another upward-facing channel within \SI{32}{ns} for a two of four majority logic requirement \cite{Barwick2017}. The thresholds are tuned such that the stations obtain a trigger rate less than \SI{10e-3}{Hz}, corresponding to a trigger threshold of approximately 4.4 times the RMS value of the ambient Gaussian-distributed thermal noise (SNR$\sim$4.4)\footnote{SNR is defined as the maximum amplitude divided by the RMS thermal noise} \cite{Barwick2017}. Since the amplifier gain is slightly temperature dependent, the exact trigger SNR threshold also varies slightly with time.

\section{Cosmic rays identification criteria}

A high purity sample of cosmic-ray candidates is needed to study the polarization resolution. Since impurities in the cosmic-ray sample highly compromise the resolution measurement, the purity of the cosmic-ray sample is here prioritized over the efficiency of the identification criteria. A set of cuts based on simulation studies and the known properties of cosmic rays were developed based on the temporal distribution of the events, correlation of the signal waveform with simulated cosmic-ray templates, signal amplitude, and the reconstructed signal arrival direction. The cuts discussed below are applied to data in the sequence in which they are discussed. The number of events passing each cut is shown in Table \ref{table:cuts}.

\subsection{Correlation analysis based on simulations}
\label{sec:MCdataset}

Simulation studies discussed in this paper are based on a set of simulated events generated using the CoREAS software. CoREAS is a Monte Carlo code for simulation of radio emission from extensive air showers \cite{CoREAS2013}. The simulated data set is a large library consisting of 1000 showers with 160 observer positions, each distributed over a \emph{star-shape} layout that efficiently samples the radio footprint. Those simulated events cover a large range of arrival directions and energies.In all studies presented in this article, the events of the MC data set are reweighted to resemble the energy and arrival direction distribution of the cosmic-ray flux.

After correcting for the amplifier response, the measured voltage traces from the four upward-facing LPDAs are correlated with the set of simulated noise-less signal templates, generated by convolving the simulated air showers with the known detector response. Averaging the correlation coefficient $\chi$ for each template, a mean normalized correlation coefficient $\overline{\chi}$ is calculated for each channel. For the two pairs of parallel upward-facing LPDAs in each station, $\overline{\chi}$ is averaged within each pair and is defined as the correlation of the pair. The larger of the $\overline{\chi}$ values obtained for each of the two antenna pairs is defined to be the final event-wise correlation $\overline{\chi}$ which can be compared with simulation.

\begin{table}
\centering
\begin{tabular}{c  c  c c} 
\hline \hline
cut name & \# events & \# events  & cut efficiency  \\
& (station32) & (station52) & (simulation) \\
\hline
rate cut & 77366  & 13171  & N/A \\
$\overline{\chi}$ cut & 126 & 104  & 0.99 \\
zenith cut & 99 & 85 &   0.91\\
SNR cut & 24   & 15 & 0.36 \\
\hline\hline
\end{tabular}
\caption{Table of cuts applied in the cosmic-ray identification process. Cut efficiency is calculated from simulation; it is defined as the number of simulated cosmic rays that pass the cut divided by the number of cosmic rays before the cut.}
\label{table:cuts}
\end{table}

\subsection{Trigger rate cut (rate cut)}

\begin{figure}[t]
\centering
  \includegraphics[scale=0.49]{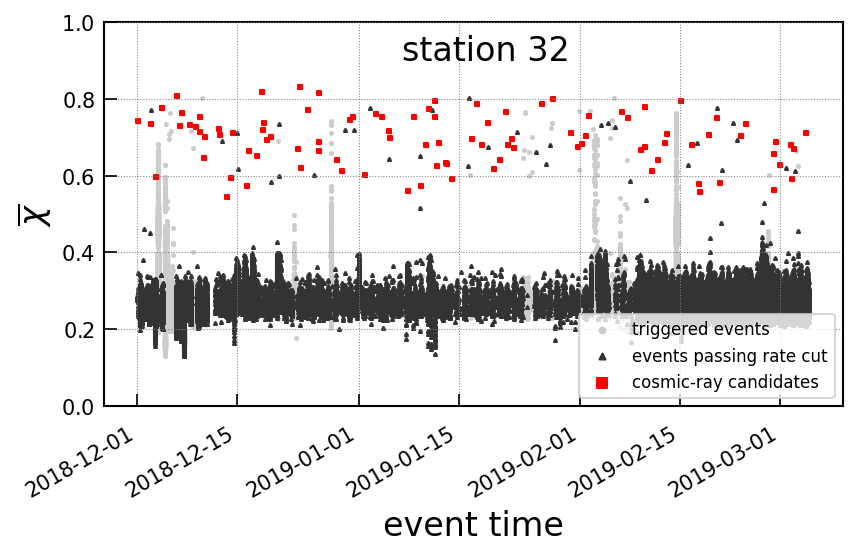}
  \includegraphics[scale=0.49]{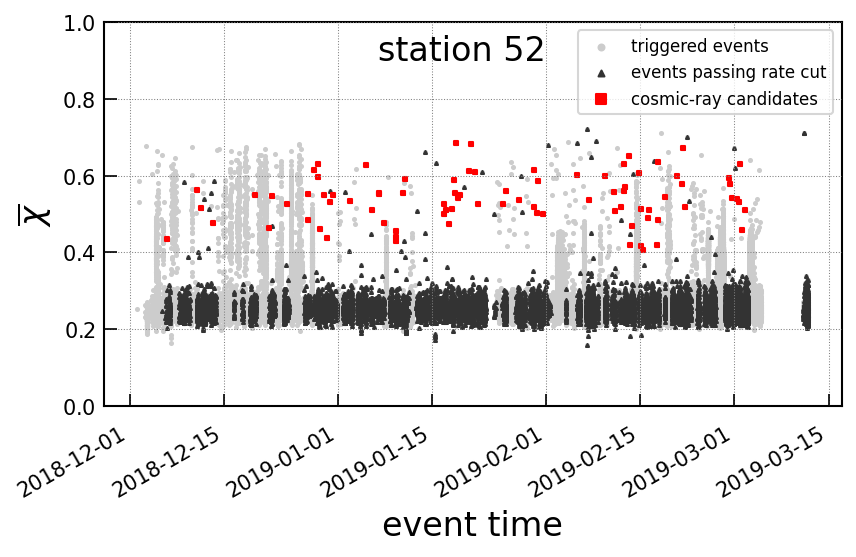}
  \caption{Temporal distribution of triggered events of the two stations. Grey dots are events that failed the trigger rate cut. Black dots are events that passed trigger rate cut but failed the $\overline{\chi}$ cut. Red squares are events that passed all cuts and are identified as cosmic-ray candidates.}
  \label{fig:time}
\end{figure}

Cosmic rays are expected to be distributed randomly in time. On the other hand, man-made signals and wind-induced radio-frequency backgrounds tend to cluster in time \cite{Barwick2017,Dave2021}. Therefore, removing temporally clustered events improves the purity of the data without significant cosmic ray efficiency loss. The trigger rate cut is designed such that temporally-clustered high-$\overline{\chi}$ events are removed. (This cut specifically targets events with high $\overline{\chi}$; low $\overline{\chi}$ events caused by thermal fluctuation are triggered at a high frequency and can be easily removed in subsequent steps.) The rate cut requires the number of triggered events with $\overline{\chi}$ greater than 0.4 to be less than 4 for each 12-hour period of data collection. If this criterion is not met, all events from the corresponding 12 hours of data collection are removed from further consideration. This threshold corresponds to a \SI{9.3e-5}{Hz} trigger rate. Based on the cosmic-ray flux calculated in previous studies, the probability that a cosmic ray is excluded by this cut is \SI{1.7e-7}{} \cite{Barwick2017}. The time distribution of all events is shown in Fig.~\ref{fig:time}; periods of high rates are evident in that plot.

\subsection{Correlation cut ($\overline{\chi}$ cut)}
\begin{figure}[t]
\centering
  \includegraphics[scale=0.5]{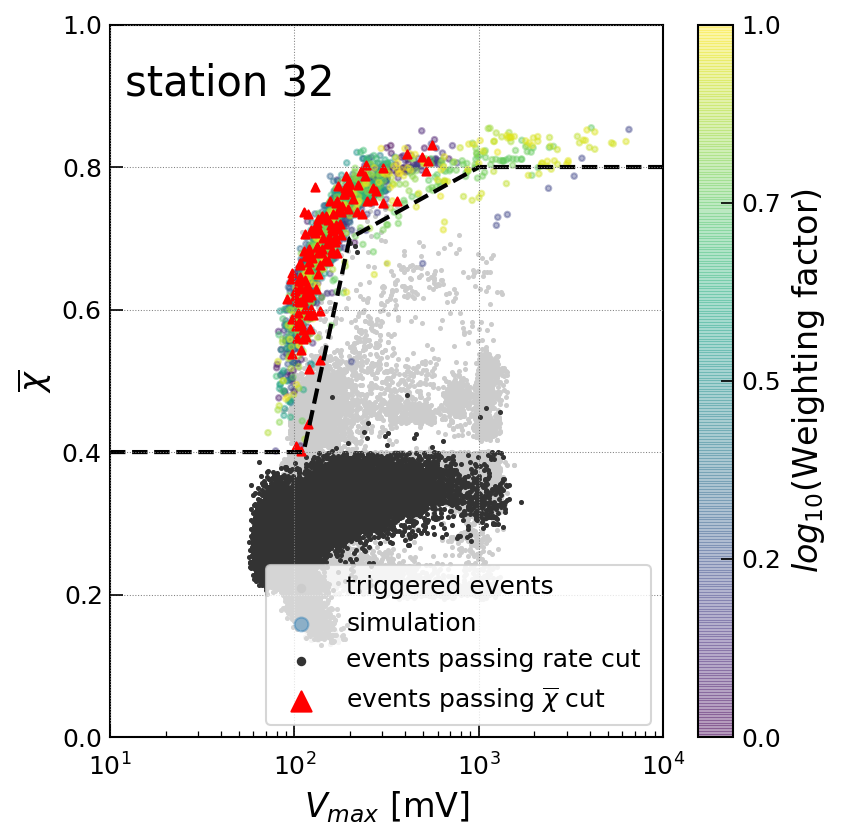}
  \includegraphics[scale=0.5]{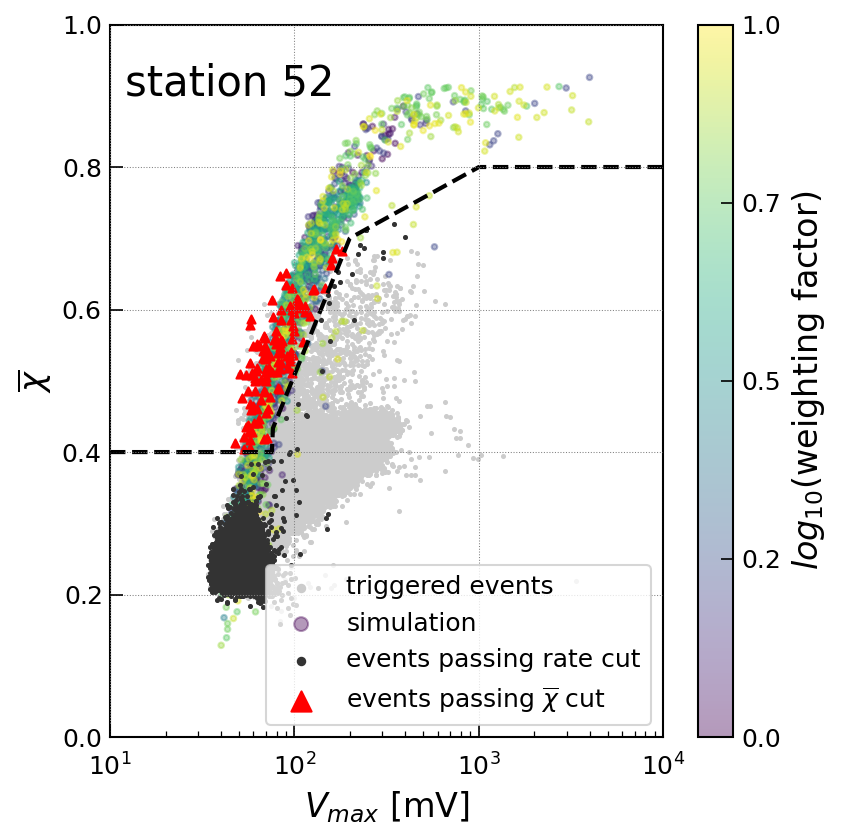}
  \caption{Comparison plot of correlation, $\overline{\chi}$ as a function of max amplitude of the event after imposing bandwidth restrictions,$V_{max}$. Grey dots are events that failed the rate cut. Black dots are events that passed rate cut but failed $\overline{\chi}$ cut. Red triangles are cosmic-ray candidates. Green dots  are simulated cosmic rays. The simulated cosmic rays are re-weighted according to the color bar shown to follow the measured cosmic-ray flux distribution.}  \label{fig:corr_amp}
\end{figure}

Over 99\% of the events passing the trigger rate cut are low-amplitude events triggered by random thermal fluctuations. Due to their thermal noise nature, these events can be easily identified and removed by correlating their waveform to the simulated cosmic-ray templates. We note that $\overline{\chi}$ values for cosmic rays are expected to be positively correlated to their amplitude, since large amplitude cosmic rays are more prominent over thermal noise and therefore have more similar waveforms to noise-less simulated templates. On the other hand, background events such as man-made radio signals or electronic noise exhibit obvious non-cosmic ray features (such as single frequency oscillations) and therefore have lower $\overline{\chi}$ values. Here, we impose a cut on the relation between $\overline{\chi}$ and the maximum amplitude $V_{\text{max}}$, with $V_{\text{max}}$ defined as the maximum amplitude of all four upward-facing channels after applying an \SI{80}{MHz} - \SI{500}{MHz} band-pass filter. In this analysis, we apply the same correlation cut as previously used in \cite{Barwick2017}. The cut threshold as a function of $V_{\text{max}}$ is shown as the dashed line in Fig.~\ref{fig:corr_amp}. Events above the cut line are retained for further analysis.

\subsection{Zenith angle cut}
A final cut is made, requiring the reconstructed zenith angle $\theta$ of the event to exceed \ang{40}. Assuming an isotropic distribution of cosmic rays, the number of incident cosmic rays with $\theta$ less than \ang{40} comprises only 23.4\% of the entire cosmic-ray population. Moreover, low-zenith cosmic rays are less likely to trigger a radio detector because of the small illuminated area on the ground \cite{Glaser_2019}. 
We performed a dedicated simulation study (using the data set described in Sec.~\ref{sec:MCdataset}) and found that only 9\% of triggered cosmic rays have $\theta$ smaller than \ang{40}.
On the other hand, wind-induced events, radio signals from airplanes, as well as other noise events whose arrival direction is not correctly reconstructed typically have low $\theta$ values. Therefore, subselecting events with $\theta$ exceeding \ang{40} rejects a large portion of background events without losing many cosmic rays.

Events passing the above cuts are identified as cosmic-ray candidates.

\subsection{SNR quality cut}

As referenced in Section \ref{sec:polarization}, an additional cut can be imposed to select high quality events. We require an SNR exceeding 4.5 in all four upward-facing channels. This quality cut can be used to ensure an accurate arrival direction reconstruction, which is a requirement for a correct polarization reconstruction. 

\section{Data set}

Two sets of cosmic-ray candidates from the two stations were identified with the criteria discussed above. In order to measure the polarization resolution with maximal statistics, it is desirable to combine the two sets of cosmic ray candidates. Here, the properties of the two set of cosmic rays are studied to make sure the two sets are compatible.

\subsection{Station 32}

127,091 events were triggered on station 32 between Dec 1 2018 and March 15 2019. The live time of the station in this data taking season is 2,541 hours. Among these triggered events, 77,366 of them passed the trigger rate cut. The dead time due to the trigger rate cut is 502 hours, which corresponds to 20\% of the total live time of the season. From the events that passed the trigger rate cut, 126 passed the correlation cut. Of these, 99 events pass the zenith cut and were identified as cosmic ray candidates, corresponding to an event rate of \SI{1.3e-5}{Hz}, or 1.2 events per day. 

\begin{table}
\centering
\begin{tabular}{c  c  c} 
\hline \hline
 & station32 & station52 \\
\hline
Total live time (hours) & 2541  & 1976.5 \\

Dead time from rate cut (hours) & 502 & 1043.5 \\

Actual live time (hours) & 2039 &933 \\

Event rate & $1.3\times10^{-5}$ Hz &$2.5\times10^{-5}$ Hz\\
& \SI{1.2 \pm 0.1}{/day} & \SI{2.2 \pm 0.2}{/day}\\
\hline \hline
\end{tabular}
\caption{Live time and event rate of the two stations.}
\label{table:rate}
\end{table}

\subsection{Station 52}
92,020 events triggered station 52 in the same data taking season, with a corresponding station live time of 1,976.5 hours. Among these triggered events, 13,171 of them passed the trigger rate cut. The dead time due to the trigger rate cut is 1043.5 hours, which corresponds to 53\% of the total live time of the season. Among events that passed the trigger rate cut, 104 of them passed the correlation cut. 85 of them passed the zenith cut and were identified as cosmic ray candidates. This corresponds to an event rate of \SI{2.5e-5}{Hz}, or 2.2 events per day. 

Live time and event rate data for the two stations are shown in Table~\ref{table:rate}.

\subsection{Comparison between the two stations}

With the same cut criteria, the two stations measured different cosmic-ray event rates in the same data taking season. Another difference between the two stations is that events from station 32 on average have higher $\overline{\chi}$ and larger $V_{\text{max}}$. Our study indicates that these differences are a direct result of differences in the hardware of the two stations, and are reproducible in simulation.

The difference in $V_{\text{max}}$ follows directly from the gain difference of the amplifiers of the two stations. As shown in Fig.~\ref{fig:amp_gains}, the gain of the station 32 amplifier is larger than the gain of the station 52 amplifier, resulting in a higher $V_{\text{max}}$ distribution for station 32.

The difference in $\overline{\chi}$ is, in part, also a result of the difference in the shape of the frequency response for the different amplifiers used for the two stations. As shown in Fig.~\ref{fig:amp_gains}, the amplifier of station 32, series-100, has a gain that peaks at \SI{100}{MHz}, which corresponds to the oscillation frequency of the tail of a cosmic-ray waveform. On the other hand, the amplifier of station 52 has a lower gain at \SI{100}{MHz}, resulting in the low frequency tail being less amplified and therefore more susceptible to noise. The low frequency tail of the waveform typically spans over half of the length of the entire waveform. As a result, the correlation of the tail dominates the correlation of the entire waveform. Since the series-100 better amplifier separates the low frequency tail from noise, station 32 events tend to correlate better to noiseless cosmic-ray templates. This hypothesis was confirmed with a simulation study. With the gain of the two amplifiers normalized to the same value so that they have the same RMS noise and with the same trigger threshold, simulated events from station 32 give an average $\overline{\chi}$ (0.63) significantly higher than events from station 52 (0.52).

\begin{figure}[t]
\centering
  \includegraphics[scale=0.5]{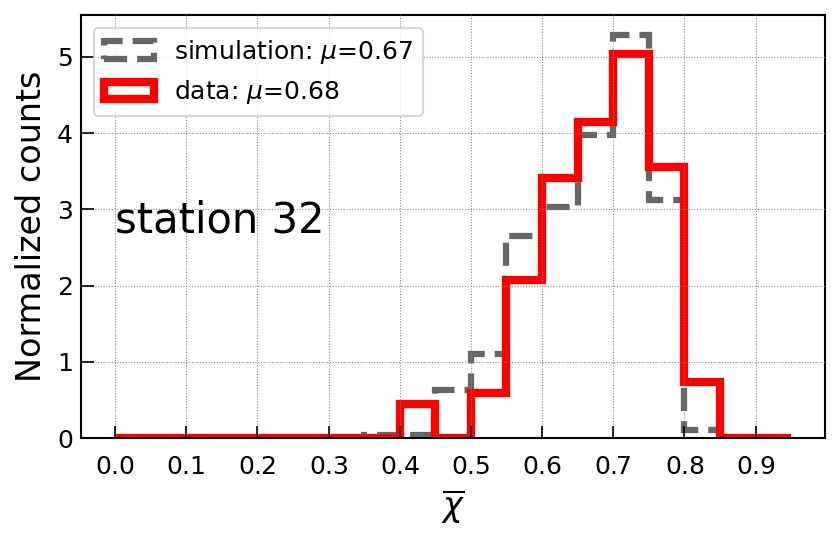}
  \includegraphics[scale=0.5]{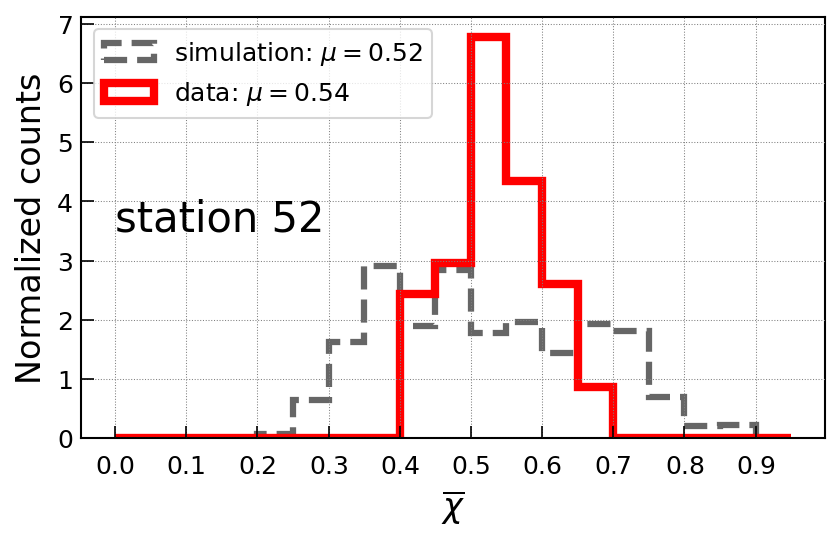}
  \caption{Distribution of $\overline{\chi}$ for data and simulated events of the two stations. Data is represented by the red histogram and simulation is represented by the grey dashed histogram. The number of events is normalized so that data and simulation are comparable. The mean of the distributions $\mu$ is also shown in the Figure.}
  \label{fig:corr}
\end{figure}

The $V_{\text{max}}$ and $\overline{\chi}$ distributions of the two stations and simulation are shown in  Fig.~\ref{fig:corr_amp} and Fig.~\ref{fig:corr}. As outlined above, the difference between the stations is reproduced in simulation.

The distributions of station 32 agree with simulation. For station 52, as shown in Fig.~\ref{fig:corr}, there is a deficiency in events with a $\overline{\chi}$ smaller than 0.4 or greater than 0.7 in data compared to simulation. The lack of events with a $\overline{\chi}$ smaller than 0.4 is a direct consequence of the minimum $\overline{\chi}$ cut. On the other hand, as shown in Fig.~\ref{fig:corr_amp}, events with $\overline{\chi}$ greater than 0.7 typically have $V_{\text{max}}$ near \SI{300}{mV}, close to the voltage at which the amplifier saturates and signals can no longer be measured reliably. Since the simulation does not account for amplifier saturation,  events are overpredicted at large $\overline{\chi}$ in simulation.

The bandwidth difference between the two amplifiers also contributes to the difference in observed event rates between the two stations. As shown in Fig.~\ref{fig:amp_gains}, the series-100 amplifier on station 32 has almost twice the bandwidth of the series-300 amplifier of station 52. For the same noise temperature, station 32 captures wide-spectrum noise up to approx. \SI{1}{GHz}, while station 52 only captures noise up to approx. \SI{500}{MHz}. Since cosmic ray signals have little power above \SI{500}{MHz}, a cosmic ray event in station 32 will have a smaller SNR compared to the same cosmic ray observed in station 52, as its noise level is higher due to the larger bandwidth. As a result, station 32 requires cosmic rays with higher energies to reach the same trigger threshold (4.4 SNR). In other words, compared to station 32, station 52 has the ability to trigger off cosmic rays with lower energies, resulting in a higher detection rate. This hypothesis is confirmed with a simulation study. When triggering off the same collection of simulated cosmic rays with thermal noise properly added, 20\% of simulated cosmic rays triggered station 52 while only 11\% triggered station 32. The simulated trigger efficiency ratio (1.82) between the two stations is identical to the measured event rate ratio in data.

With an understanding of the difference between the performance of the two stations in hand, we now combine the data from the two stations to study polarization.

\section{Polarization reconstruction}
\label{sec:polarization}

\begin{figure}[t]
\centering
  \includegraphics[scale=0.75]{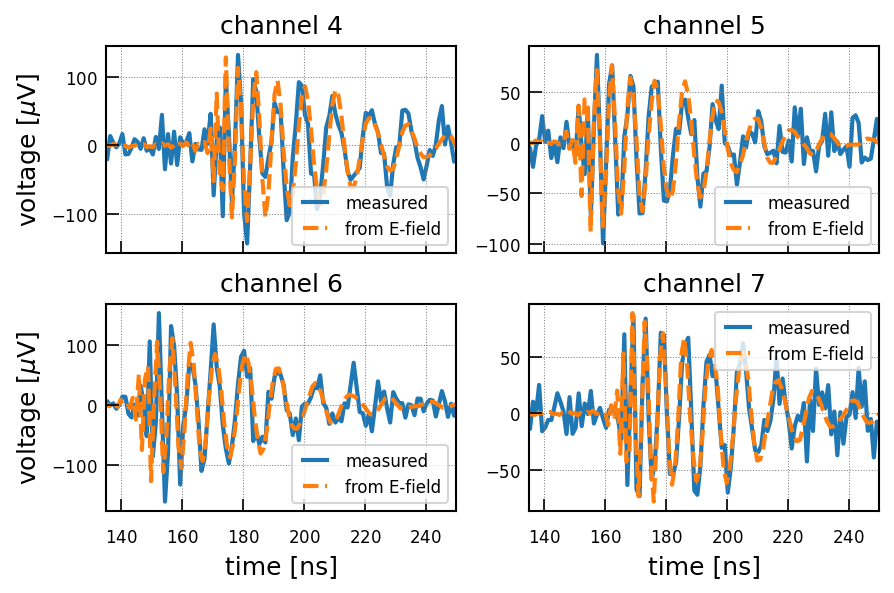}
  \caption{An example of electric field reconstruction using the forward-folding technique. The blue curves represent the measured voltage traces and the orange dashed curves are the analytic solution of the forward-folding technique.}
  \label{fig:forward_folding}
\end{figure}

In order to reconstruct the polarization of an air shower, the electric field needs to be reconstructed. In this study, the forward-folding technique is used to reconstruct the electric field. Instead of numerically recovering the incident electric field (frequency bin by frequency bin), the forward-folding technique fits an analytic model of the electric-field pulse directly to the measured voltages in the time domain. This technique is found to have significantly better accuracy compared to standard unfolding methods, especially for small SNR events \cite{NuRadioReco}. As shown in Fig.~\ref{fig:forward_folding}, the measured voltage traces agree with the analytic solution of the forward-folding technique to high precision. The agreement is particularly good for the low frequency components.

The polarization angle is defined to be the angle between the  $\textbf{e}_{\phi}$ and $\textbf{e}_{\theta}$ components of the electric field vector, where $\textbf{e}_{\phi}$ is the s-polarized radiation direction of the electric field and $\textbf{e}_{\theta}$ is the p-polarized radiation direction of the electric field \cite{Barwick2017}. The polarization is calculated using $P_{\text{rec}}=\arctan(\sqrt{E_{\phi}}/\sqrt{E_{\theta}}$), where $E_{\phi}$ and $E_{\theta}$ are the energy fluences of the $\phi$ and $\theta$ components of the reconstructed electric field, respectively. Here, the energy fluence is used to calculate the polarization angle in order to stay consistent with previous studies \cite{NellesICRC2019,ARIANNA2020Polarization}. The requirement that all upward-facing channels have an SNR larger than 4.5 ensures a precise arrival direction reconstruction, which is needed to for an accurate signal polarization reconstruction\cite{NuRadioReco}. This cut selects 39 events out of the 184 cosmic ray candidates in the combined data set. It should be noted that a similar cut is not required for neutrino events, since more advanced reconstruction methods can be implemented (e.g.\ using a forward-folding technique or a neural network-based reconstruction) that achieve a precise signal arrival direction reconstruction for neutrino events without the requirement of large SNR in all downward-facing LPDAs \cite{GaswintPhD, ICRC2021DLdirection,RNO-G:2021zfm,ARIANNADirectionICRC2021}.These techniques can also be implemented for cosmic-ray reconstruction in the future.

\subsection{Calculation of predicted polarization}
In order to evaluate the performance of the reconstruction, a prediction of polarization is needed. The polarization characteristics of air showers are dominated by the geomagnetic effect, with a small contribution from the Askaryan effect \cite{CoREAS2013,AERAPolarization,Glaser_2019}. Therefore, a relatively accurate prediction of polarization can be made using the geomagnetic polarization $P_{\text{exp}}=\vec{v}\times\vec{B}$, where $\vec{v}$ is the signal arrival direction and $\vec{B}$ is the local geomagnetic field vector. This prediction accounts for the dominant geomagnetic effect but not for the subdominant Askaryan effect. We define the polarization uncertainty $\Delta P$ as $|P_{\text{rec}}-P_{\text{exp}}|$; the 68th percentile of $\Delta P$ is defined to be the polarization resolution.

\subsection{Polarization uncertainty distribution}

\begin{figure}[t]
\centering
  \includegraphics[scale=0.7]{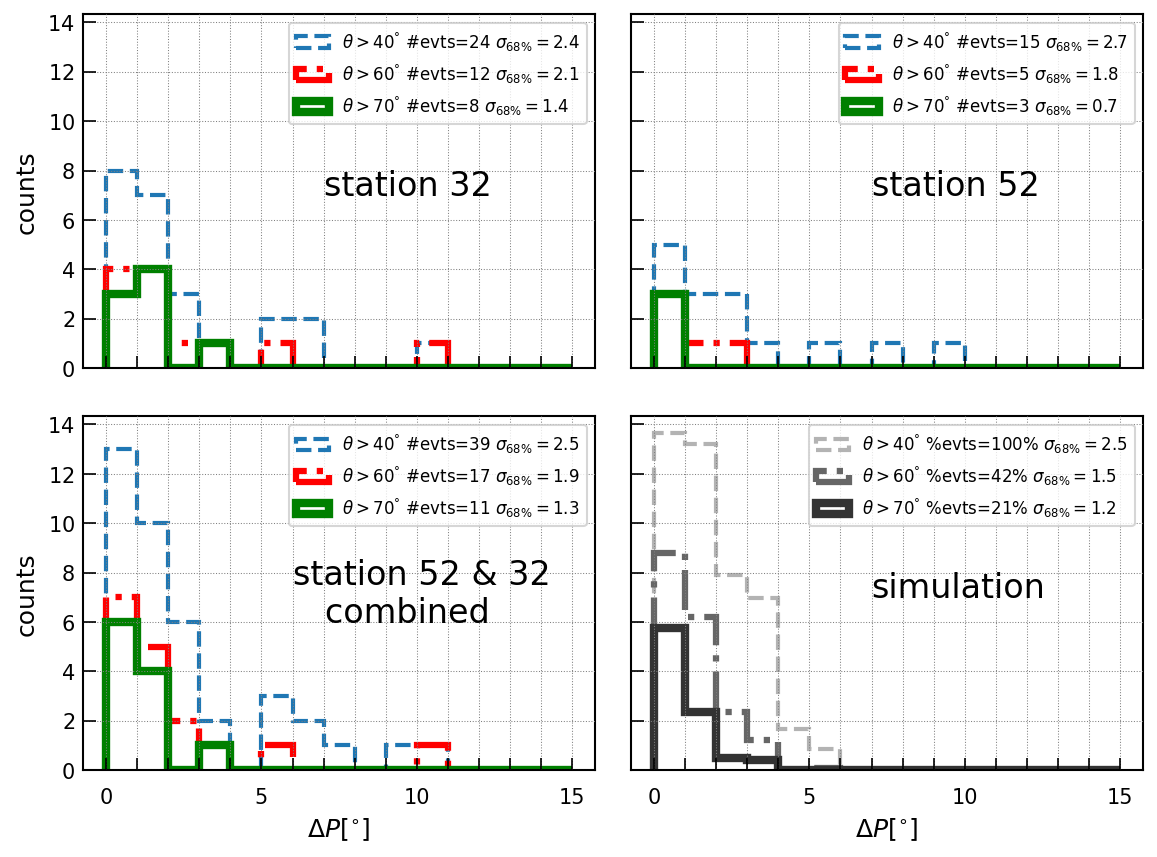}
  \caption{The distribution of $\Delta P$ of data and simulation. $\sigma_{68\%}$ represents the 68th percentile of the distribution. The number of events in simulation is normalized so that it is comparable to data. The percentage number in the simulation plot represents the ratio between the number of events above a certain $\theta$ threshold and the number of events with $\theta$ greater than \ang{40}. }
  \label{fig:pol_err}
\end{figure}

Based on the reconstructed arrival direction, the signal polarization can now be predicted and compared to the reconstructed polarization. This allows us to assess the polarization resolution directly from measured cosmic rays. The resulting distribution of $\Delta P$ is shown in Fig.~\ref{fig:pol_err}. Combining data from the two stations, the polarization resolution is measured to be \ang{2.5}.

The same reconstruction and analysis are performed on a simulated data set consisting of CoREAS simulations with the simulated events re-weighted so that they correspond to the expected cosmic-ray flux. Each shower was re-used several times with random core positions. The simulated observer position closest to the ARIANNA station was selected for a full detector simulation, including the addition of thermal noise obtained from a library of periodic detector readouts. At this stage, the the signal polarization in simulated events can be reconstructed in the same way as data. A polarization resolution of \ang{2.5} is found for simulations, in agreement with data. 

\begin{figure}[t]
\centering
  \includegraphics[scale=0.8]{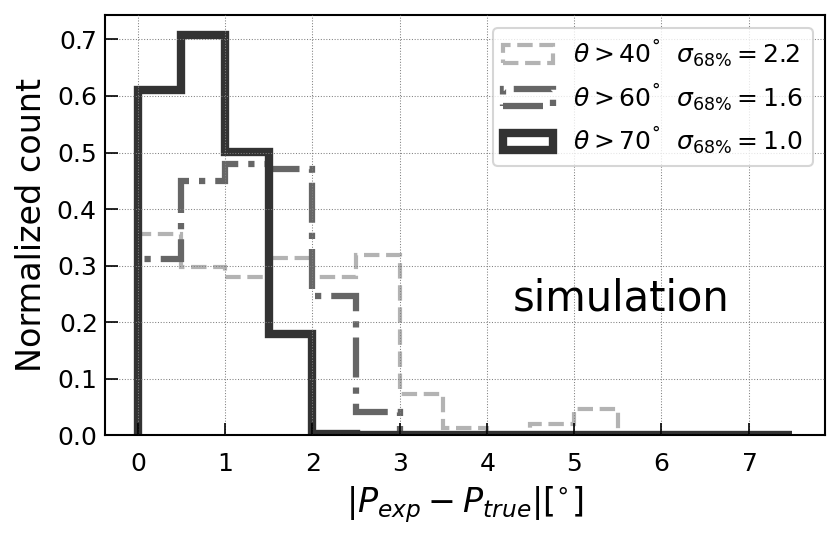}
  \caption{Distribution of the difference between the true polarization (directly obtained from CoREAS simulations) and the expected geomagnetic polarization calculated from the air-shower direction $| P_{\text{true}} - P_{\text{exp}}|$ of simulated events. $\sigma_{68\%}$ represents the 68th percentile of the distribution.}
  \label{exp_true}
\end{figure}

\subsection{Source of polarization uncertainty}
We performed a simulation study to determine the uncertainty that is introduced by using in the $\Delta P$ calculation the geomagnetic signal polarization instead of the true signal polarization where also the Askaryan effect contributes to. We find that the measured polarization uncertainty $\Delta P$ is dominated by uncertainties in the expected polarization $P_{\text{exp}}$ rather than the reconstructed polarization $P_{\text{rec}}$. In our calculation, $P_{\text{exp}}$ only accounts for the contribution from the geomagnetic emission but not the sub-dominant Askaryan emission. However, the contribution of the Askaryan effect is not negligible given the precision of our current measurements. The difference between the true polarization $P_{\text{true}}$ and the geomagnetic polarization expectation $P_{\text{exp}}$ is shown in Fig.~\ref{exp_true}, where $P_{\text{true}}$ is the ground truth polarization of the simulated events. As before, the library of CoREAS simulations is reweighted to the cosmic-ray flux. Comparing Fig.~\ref{exp_true} to the $\Delta P$ distribution of simulated events in Fig.~\ref{fig:pol_err}, it can be seen that a large part of $\Delta P$ originates from the difference between $P_{\text{true}}$ and $P_{\text{exp}}$.

In order for $P_{\text{exp}}$ to accurately represent $P_{\text{true}}$, it is desirable to subselect events in which the Askaryan effect is minimized. The influence of the Askaryan effect on polarization is related to the station's position relative to the air shower axis. It is also inversely correlated to the angle between $\vec{B}$ and $\vec{v}$, which reduces the strength of the geomagnetic emission. Since $\vec{B}$ is nearly vertical at Ross-Ice-Shelf, the dependence on this angle is dominated by the dependence on $\theta$. Since the strength of the geomagnetic emission depends on the air density column in which the shower develops (increasing with decreasing air density \cite{GlaserErad2016}), the influence of the Askaryan contribution is minimized at large $\theta$.  We independently confirm this prediction using our library of CoREAS simulations where we find that the $P_{\text{true}}$, $P_{\text{exp}}$ difference decreases with increasing zenith angle $\theta$ (cf. Fig.~\ref{exp_true}).
Therefore, $P_{\text{exp}}$ best approximates the true polarization at large $\theta$. In order to evaluate the true polarization resolution of the detectors, the analysis is therefore repeated on the subset of events having large $\theta$.

In addition,the SNR of cosmic-ray candidates does not change significantly with increasing $\theta$, with the mean SNR of cosmic-ray candidates passing the SNR quality cut with $\theta$ above \ang{40} being 11.0 and the mean SNR of cosmic-ray candidates passing the SNR quality cut with $\theta$ above \ang{70} being 11.6. The independence of SNR on $\theta$ ensures that the subselection of large $\theta$ events does not bias the calculation of $\Delta P$ towards large SNR events.

As shown in Fig.~\ref{fig:pol_err}, $\Delta P$ is significantly reduced when subselecting events above a certain $\theta$ threshold. As mentioned previously, we attribute this to the decreasing contribution to the polarization from the Askaryan emission as $\theta$ increases (for the geomagnetic field at the Ross-Ice-Shelf). With limited statistics, a resolution of \ang{1.3} is obtained when we subselect events with $\theta$ greater than \ang{70}. This agrees with the \ang{1.2} resolution obtained from simulation.

Another contributor to $\Delta P$ is the systematic uncertainties in the antenna response measurements, but the amplitude of the contribution to $\Delta P$ is currently not known from first principle calculations. Therefore, the measured polarization resolution of \ang{1.3} obtained from the $\Delta P$ distribution at large $\theta$ provides an upper limit to the unknown systematic uncertainties of the antenna response. The width of the $\Delta P$ distribution can be reduced in the future if the systematics are more accurately understood and then corrected in the analysis.

\subsection{Data from the 2017-2018 season}

\begin{table}
\centering
\begin{tabular}{c  c  c} 
\hline \hline
 &  polarization resolution($^{\circ}$) &  polarization resolution($^{\circ}$) \\
& (2017-2018) & (2018-2019)\\
\hline
station 32 & 6.9  & 2.4 \\
station 52 & 5.5 & 2.7 \\
\hline\hline
\end{tabular}
\caption{Comparison of polarization resolution between  season 2017-2018 and 2018-2019. Numbers in the table are the 68th percentile of $\Delta P$ of cosmic-ray candidates passing the SNR quality cut.}
\label{table:2018}
\end{table}

Data from the 2017-2018 season was analyzed with the same cosmic-ray identification criteria and polarization reconstruction procedures. For station 32, 83 cosmic ray candidates were identified from 26,216 triggered events. A polarization resolution of \ang{6.9} was measured for these events, which agrees with the previous result of \ang{7.0} using the same data set \cite{NellesICRC2019}. Subselecting events with $\theta$ greater than \ang{70}, a polarization resolution of \ang{3.0} was measured.

For station 52, 97 cosmic ray candidates were identified from 83,600 triggered events. A polarization resolution of \ang{5.5} was measured for these events. Subselecting events with a $\theta$ greater than \ang{70}, a polarization resolution of \ang{3.3} was measured.

As shown in Table~\ref{table:2018}, compared to data from 2018-2019, data from 2017-2018 yielded significantly worse polarization resolution. We attribute this to the shallow deployment of the stations in the 2017-2018 season. According to snow accumulation records, the antennas of the two stations in the 2017-2018 season were \SI{1.10}{m} shallower in the snow compared to 2018-2019. The signal propagation model, as well as the modelling of the antennas, assume the antennas are surrounded by uniform, infinite snow. However, in the 2017-2018 season, the two stations were too close to the snow surface for this assumption to be valid. We therefore believe the measurement with data from 2018-2019 to be a more accurate determination of the polarization resolution of the detectors. Further studies are needed to confirm this hypothesis.

\subsection{Comparison to previous results}

Compared to the work in \cite{NellesICRC2019}, we measure considerably improved polarization resolution. One of the main reasons for this improvement is the improved data set. The previous study \cite{NellesICRC2019} used data from 2017-2018, when the antennas were so close to the snow surface that the signal propagation model was less reliable. After a year of snow accumulation, in 2018-2019, the stations were at a depth where the model better matches the actual signal propagation properties, yielding significantly improved reconstruction performance. 

Another contributing factor to the enhancement in polarization resolution is a higher purity data sample. Compared to the cuts used in \cite{NellesICRC2019}, an additional zenith cut is applied in this analysis. Also, compared to the trigger rate cut in \cite{NellesICRC2019}, the trigger rate cut in this paper excludes background events more effectively. The previous trigger rate cut focused on the event rate over a short period of time (1 hour) \cite{Barwick2017}. However, wind storms typically last for multiple hours or days. The trigger rate cut in this paper focuses on the trigger rate over a longer period of time (12 hours) and removes windy periods more effectively. As a result, a larger fraction of wind-induced events, which have similar waveforms to cosmic-ray events, are removed in this study, leading to a purer data set.

\section{Conclusion}
We have presented a measurement of the polarization resolution using radio signals from cosmic-ray air showers triggering the ARIANNA high-energy neutrino detector. 
184 cosmic ray candidates were identified from a 3-month data-taking period between Dec 2018 to Mar 2019 from two detector stations. 39 events that have an SNR greater than 4.5 in all upward channels were selected for the polarization study. We found that $P_{\text{rec}}$ agrees within \ang{2.5} with $P_{\text{exp}}$, the expected polarization obtained solely based on the geomagnetic emission and the reconstructed air-shower direction. Moreover, we found that the uncertainty in $P_{\text{exp}}$ arises primarily from the subdominant Askaryan radio emission in air showers, not accounted for in the $P_{\text{exp}}$ numerical extraction. Subselecting events with larger $\theta$, where the contribution from the Askaryan emission is smallest, we found an improved agreement between $P_{\text{rec}}$ and $P_{\text{exp}}$, which also agreed with simulation. In particular, the inferred resolution improved to \ang{1.3} when subselecting events with zenith angles larger than \ang{70}.

We have furthermore found that a (too) shallow deployment of the LPDAs detrimentally affects the polarization reconstruction. This can be addressed by deploying the LPDAs of future detectors at least a meter below the snow surface. 

These results are consistent with {\it in situ} pulsing measurements, where a polarization reconstruction precision of \ang{1} has been found with a slowly-varying systematic variation with an RMS error of \ang{2.7}, as a function of emitter depth. We note that the pulsed measurements are subject to a somewhat different set of systematic uncertainties \cite{ARIANNA2020Polarization} than those described herein. 

As the radio signals expected from neutrino interactions in ice are similar to the air-shower signals, this measurement constitutes an important in-situ test for the capability to reconstruct the neutrino direction with shallow in-ice detectors, for which polarization reconstruction is the dominant uncertainty. This technique of using cosmic rays as a calibration tool for neutrino radio detectors can be adapted to other sites with minor modifications. These results are therefore an important verification test for the planned IceCube-Gen2 \cite{Gen2WhitePaper,Gen2RadioICRC}.

\section{Acknowledgement}
We are grateful to the U.S. National Science Foundation-Office of Polar Programs, the U.S. National Science Foundation-Physics Division (grant NSF-1607719) for supporting the ARIANNA array at Moore's Bay, and NSF grant NRT 1633631. Without the invaluable contributions of the people at McMurdo Station, the ARIANNA stations would have never been built. We acknowledge funding from the German research foundation (DFG) under grants GL 914/1-1 and NE 2031/2-1, the Taiwan Ministry of Science and Technology, the Swedish Government strategic program Stand Up for Energy, MEPhI Academic Excellence Project (Contract No.  02.a03.21.0005) and the Megagrant 2013 program of Russia, via agreement 14.12.31.0006 from 24.06.2013. The computations and data handling were supported by resources provided by the Swedish National Infrastructure for
Computing (SNIC) at UPPMAX partially funded by the Swedish
Research Council through grant agreement no. 2018-05973.

\bibliographystyle{JHEP}
\bibliography{bib}

\end{document}